\documentclass[aps]{revtex4}
\usepackage{amsfonts}
\usepackage{amsmath}
\usepackage{amssymb}
\usepackage{epsfig}
\begin{document}
\title[Sensitivity and specificity of DNA chips]{Sensitivity, Specificity 
and the Hybridization Isotherms of DNA Chips}
\author{A. Halperin}
\affiliation{UMR 5819 (UJF, CNRS, CEA) DRFMC/SI3M, CEA Grenoble, 
17 rue des Martyrs, 38054 Grenoble cedex 9, France}
\author{A. Buhot}
\affiliation{UMR 5819 (UJF, CNRS, CEA) DRFMC/SI3M, CEA Grenoble, 
17 rue des Martyrs, 38054 Grenoble cedex 9, France}
\author{E. B. Zhulina}
\affiliation{Institute of Macromolecular Compounds of the Russian Academy 
of Sciences, 199004 St Petersburg, Russia}


\pacs{PACS number}


\begin{abstract}
Competitve hybridization, at the surface and in the bulk, lowers the sensitivity 
of DNA chips. Competitive surface hybridization occurs when different targets
can hybridize with the same probe. Competitive bulk hybridization takes place 
when the targets can hybridize with free complementary chains in the solution. 
The effects of competitive hybridization on the thermodynamically attainable 
performance of DNA chips are quantified in terms of the hybridization
isotherms of the spots. These relate the equilibrium degree of the hybridization
to the bulk composition. The hybridization isotherm emerges as a Langmuir 
isotherm modified for electrostatic interactions within the probe layer. 
The sensitivity of the assay in equilibrium is directly related to the slope 
of the isotherm. A simpler description is possible in terms of $c_{50}$s 
specifying the bulk composition corresponding to $50\%$ hybridization 
at the surface. The effects of competitive hybridization are important for the 
quantitative analysis of DNA chip results especially when used to study point
mutations.
\end{abstract}

\maketitle

\section{Introduction}

DNA microarrays allow to interrogate the base sequence of DNA or RNA
chains. They can be used to detect pathogens, identify genetic defects,
monitor gene expression \emph{etc.} (Marshal and Hodgson, 1998; Graves, 1999; 
Niemeyer and Blohm, 1999; Southern et al., 1999; Wang, 2000; Pirrung, 2002). 
In spite of the intense activity in this field, theoretical aspects of the 
function of DNA microarrays received relatively little attention. Early 
theoretical work focused on the dynamics of hybridization at the surface 
(Chan et al., 1995; Livshits and Mirzabekov, 1996). Recently, theoretical 
investigations considered the equilibrium hybridization isotherms of DNA chips 
(Vainrub and Pettitt, 2002; Vainrub and Pettitt, 2003) and polyelectrolyte 
aspects of the systems (Crozier and Stevens, 2003). In the following we 
present a theoretical analysis of the effect of competition between different 
possible hybridization reactions on the sensitivity and specificity of DNA chips. 
The discussion utilizes hybridization isotherms relating the equilibrium fraction 
of hybridized chains at the surface, $x$, to the composition of the bulk. The
effects are revealed by comparison of the hybridization isotherms for
competition-free situations with those obtained when competitive hybridization
is significant. They are quantified in terms of various $c_{50}$s specifying
the bulk composition corresponding to $50\%$ hybridization at the surface. A
key ingredient of our discussion is the derivation of the competition-free
isotherm as a Langmuir adsorption isotherm modified to allow for electrostatic
interactions. Our model is related to an earlier model proposed by Vainrub and
Pettitt (VP) in that both assume uniform smearing of the electrical charge of
the probe layer.

The elementary units of DNA microarrays are ``spots'' containing numerous 
single stranded DNA (ssDNA) chains, of identical sequence,
terminally anchored to the support surface. The spots are placed in a
checkered pattern so that each sequence is allocated a unique site. These
chains, or probes, preferentially hybridize with free ssDNA chains having a
complementary sequence. The microarray is immersed in a solution containing
labeled ssDNA chains whose sequence is not known and are commonly referred 
to as ``targets''. The presence of specific sequences is signalled by
hybridization on the corresponding spot as monitored by correlating the
strength of the label signal with the position of the spot (Graves, 1999).
Recently, label free detection methods, involving optical and mass sensitive
techniques, attract growing attention (Niemeyer and Blohm, 1999). These allow
to monitor the kinetics of hybridization. However, such methods measure the
total hybridization of a particular probe irrespective of the identity of the
partner. In marked contrast, selective labeling of a particular sequence
monitors only the hybridization of this target and does not report on the
hybridization of other moieties.

The unitization of DNA chips as analytical method involves immersing the
device in a solution containing a mixture of DNA chains of different sequences
and concentrations. Under such conditions, it is necessary to allow for the
role of competitive hybridization. It is useful to distinguish between two
types of competitive hybridization. Competitive surface hybridization occurs
when a number of different targets can hybridize with the same probe. Thus, a
site occupied by certain probes will preferentially hybridize DNA targets with
a perfectly matched complementary sequence. However, it will also hybridize a
certain fraction of mismatched sequences. As we shall discuss, this fraction
depends on the binding constants as well as the concentrations of the moieties
involved. Competitive hybridization at the surface clearly lowers both the
sensitivity and the specificity of the assay. When the surface competition is
significant, labeled and unlabeled detection may yield different results. No
difference is expected when all targets are labeled, as is the case when PCR
amplification is used. On the other hand, when selective labeling of specific 
targets is possible, the two techniques measure different
quantities corresponding to different isotherms. Competitive bulk
hybridization reduces the concentration of non-hybridized targets that are
available for hybridization with the probe. This takes place when the solution 
contains complementary sequences that can
hybridize with the target in the solution. Such sequences may occur either in
the same chain, leading to hairpin formation, or in different sequences
leading to interchain hybridization. Competitive bulk hybridization diminishes
thus the sensitivity of DNA chips. Its importance varies, again, with the 
binding constants and the concentrations. The issues discussed above assume 
their clearest form when DNA chips are used to identify single nucleotide
polymorphism or point mutations (Lopez-Crapez et al., 2001). 
In these situations, the DNA chip is
exposed to a mixture of targets differing from each other only in the identity
of one particular base. The fraction of the different forms is then deduced
from the relative intensity of the signals of the four spots corresponding to
the four possible sequences.

\begin{figure}
\centerline{\epsfig{file=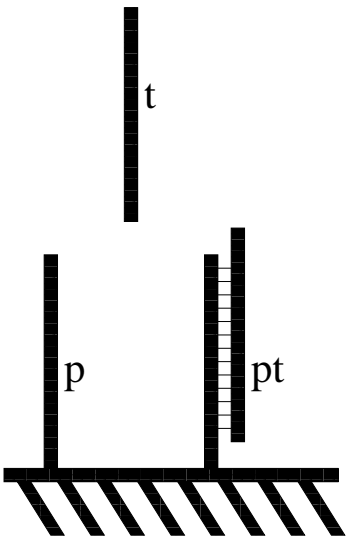,width=1.8in}}
\caption{A schematic representation of the competition free case where the
probes, $p$, can hybridize with a single target species, $t$.}
\end{figure}

\begin{figure}
\centerline{\epsfig{file=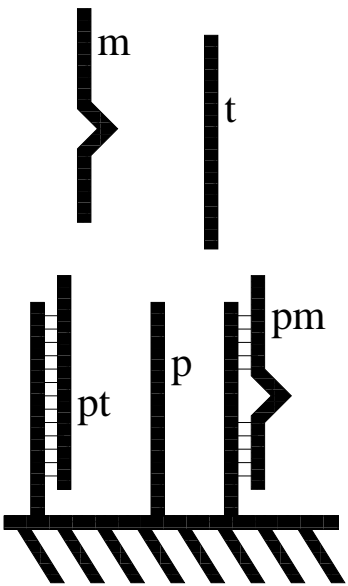,width=1.8in}}
\caption{In the competitive surface hybridization case the probes, $p$, can
hybridize with a perfectly matched target species, $t$, as well as with a
mismatched target, $m$.}
\end{figure}

\begin{figure}
\centerline{\epsfig{file=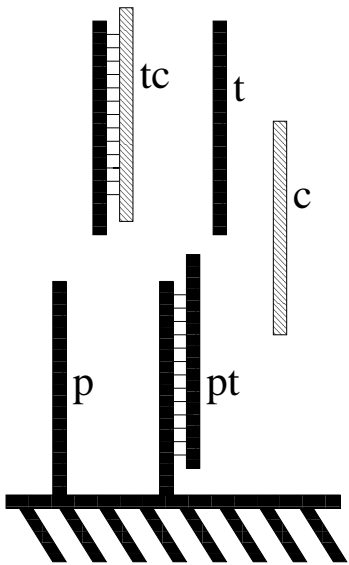,width=1.8in}}
\caption{Competitive bulk hybridization when the probes, $p$, can hybridize
with a single perfectly matched target species, $t$ but $t$ can also hybridize
in the bulk with a complementary chain, $c$. $c$ can not hybridize with $p$.}
\end{figure}

In practice, the DNA chips are immersed in the target solution for a
relatively short time. As a result, the attainment of equilibrium is not
guaranteed and rates of the different hybridization reactions play an
important role. Yet, full analysis of the reaction kinetics requires knowledge
of the equilibrium state. An understanding of the equilibrium state is also
necessary in order to identify the relative importance of kinetic and
thermodynamic controls of the performance of the DNA microarrays. Finally,
emerging evidence (Bhanot et al., 2003) suggests that the performance of DNA
chips, as measured by the number of ``false positives'', is
best at the thermodynamic equilibrium. With this in mind, we investigate the
equilibrium hybridization isotherms for three idealized but experimentally
attainable situations. These situations involve a DNA array immersed in
solutions of different composition: (i) A solution containing one species
of sigle stranded target (Figure 1). 
(ii) A solution containing two different targets that do not
hybridize in the bulk but are both capable of hybridizing with the same probe
(Figure 2) and (iii) A solution containing two different chains, a target and
a complementary chain capable of hybridizing with it in the bulk but incapable
of hybridizing with the probe (Figure 3). In all cases, we consider the case
of probes and targets of equal length i.e., the number of bases, $N$, in the
chains are identical. For brevity our discussion focuses on systems where
the hybridization at the surface has a negligible effect on the concentration
of targets in the bulk. This case corresponds to small spots or to elevated 
target concentration. 

The first two sections summarize the necessary background information for the
subsequent discussion. Thus, section II recalls the definitions of sensitivity
and other measures of the performance of analytical assays. The relationship
between sensitivity and the equilibrium hybridization isotherm is also
discussed. The relevant structural characteristics of DNA chips and important
length scales in the problem are summarized in section III. The next section,
IV, is devoted to the derivation of the competition-free hybridization
isotherm as a Langmuir isotherm modified to allow for electrostatic
interactions. Initially we obtain the hybridization isotherm for an arbitrary
electrostatic free energy density of the probe layer, $\gamma_{el}$. We then
consider the hybridization isotherms for particular functional forms of
$\gamma_{el}$ assuming a laterally uniform smearing of the electric charge. We
mostly focus on the ``diffuse layer'' model where the charge is
uniformly smeared within the probe layer thus allowing for its thickness. 
It is important to note that some of our results are actually independent 
of the model specifying $\gamma_{el}$. We conclude this section with a 
discussion of relevant experimental results and a comparison between our 
approach and the VP model. 
In the remaining sections we pursue two complementary goals: The
modifications of the hybridization isotherms to allow for competitive
hybridization and the resulting effects on the sensitivity and specificity of
the assay. Three situations are considered. The competition free case, when
the probes are exposed to a single target, is discussed in section V. This
yields upper bounds for the sensitivity and the specificity. Competitive
surface hybridization is analyzed in section VI and competitive bulk
hybridization is considered in section VII. The detailed derivation of
$\gamma_{el}$ within the diffuse layer model is described in Appendix A.
The hybridization isotherm for low salt solutions is discussed in Appendix B.

\section{On Sensitivity and the Hybridization Isotherm}

As we shall see, the equilibrium hybridization isotherms naturally suggest
characterization of the sensitivity of the assay in terms of appropriate
$c_{50}$s. This characteristic is closely related to the common definitions of
the sensitivity of analytical techniques. It is thus useful to first summarize
these definitions and their relationship to the hybridization isotherms.
Different definitions of sensitivity are available (Pardue, 1997; Ekins and
Edwards, 1997; references cited in Pardue, 1997). The IUPAC definition
identifies the sensitivity, $S_{e}$, with the slope of the calibration curve.
The calibration curve describes the measured response, $R$, to a target
concentration, $c_{t}$, $R(c_{t})$ and
\begin{equation}
S_{e}=dR/dc_{t}. \label{01}
\end{equation}
The quantitative resolution of the assay, $\Delta c_{t}$, is then specified by
\begin{equation}
\Delta c_{t}=\epsilon_{r}(c_{t})/S_e(c_{t}) \label{02}
\end{equation}
where $\epsilon_{r}$ is the measurement error as given by its standard
deviation. The detection limit, the lowest detectable $c_{t}$, is determined by
$\Delta c_{t}(c_{t}=0)$ since when the concentration $c_{t}$ is lower than
$\Delta c_{t}(c_{t}=0)$ the error is larger than the signal. The
IFCC convention identifies the sensitivity with the detection limit.

Our goal is to relate the sensitivity of DNA chips to their hybridization
isotherms. With this in mind, it is convenient to adopt the IUPAC definition.
This choice is motivated by the following observations: (i) the calibration
curve in equilibrium is closely related to the hybridization isotherm; (ii)
The measurement error depends on the measurement technique and on instrumental
characteristics. In distinction to $R(c_{t}),\epsilon_{r}$ is not related to
the calibration curve and (iii) $S_{e}$ as given by Eq.~\ref{01} plays a role in
the determination of both the qualitative resolution and the detection limit.

In the following we will assume that $R(c_{t})$ is proportional to the
equilibrium hybridization fraction at the surface, $x$ i.e., $R(c_{t})=\kappa
x+ const$ where $\kappa$ is a constant. This assumption is justified when the
following conditions are fulfilled: (i) Non-specific adsorption is negligible
and $R$ is due only to hybridization at the surface; (ii) The duration of the
experiment is sufficiently long to allow the hybridization to reach
equilibrium and (iii) the measured signal depends linearly on the amount of
oligonucleotides at the surface. It is useful to note the following points
concerning the attainability of these conditions. First, surface treatments
repressing non-specific adsorption are available for certain substrates (Steel
et al., 2000 and references cited in Steel et al., 2000). Second, the attainment
of stationary state for the hybridization may require long periods of up to 14
hours (Peterson et al., 2001; Peterson et al., 2002; Bhanot et al., 2003). 
Furthermore, the degree of hybridization may depend on the thermal history 
(heating of the substrate or the solution). In this context it is important 
to stress that, by definition, a state of thermodynamic equilibrium is both 
stationary in time and independent of the path i.e., preparation method. 
Finally, the linear range varies with the measurement technique. For example, 
when using fluorescent labels the linear regime occurs at low enough 
concentration when self-quenching is negligible (Lakowicz, 1999).

\section{Relevant Molecular Dimensions and Length Scales}

Two groups of length scales play an important role in our subsequent
discussion. One group describes the structural features of the probe layer.
The second characterizes the electrostatic interactions and their screening.
Expression of the free energies in terms of these length scales allows for
a compact formulation and the identification of the relevant dimensionless 
variables.

The structural features of the layer are determined mostly by the dimensions
of the hybridized and unhybridized probes as well as the grafting density
(Graves, 1999; Southern et al., 1999; Pirrung, 2002). The number of monomers,
nucleotides, per probe, $N$, varies over a wide range. Values of $10 \leq
N \leq 30$ are common but much higher values, of $N \approx 1000$ are attainable.
In the following we will consider systems comprising of probes and targets of
equal size in the range $10 \leq N \leq 30$. Double stranded DNA (dsDNA) is a
semiflexible chain with a persistence length $\approx 10^{3} \mathring{A}$ 
(Cantor and Shimmell, 1980). Thus, in our $N$ range double stranded 
oligonucleotides may be viewed as rigid rods with the radius of a dsDNA, 
$r=9.5 \mathring{A}$ and a projected length per monomer along the axis of 
$2b=3.4 \mathring{A}$. The corresponding parameters for ssDNA are not yet 
established. Stacking interactions between the hydrophobic bases tend to produce 
a stiff ``single stranded helix'' (Cantor and Shimmell, 1980; Bloomfield et al.,
2000; Korolev et al., 1998 and refrences cited in Korolev et al., 1998). Since
these interactions are non-cooperative this tendency is especially marked in
short ssDNA considered by us. Theoretical studies of the melting behavior of
free DNA in the bulk suggest that ssDNA can be modeled as a rigid rod with
projected length per monomer of $a \approx3.4 \mathring{A}$ and a radius of
$r_{ss} \approx 7 \mathring{A}$ (Frank-Kamenetskii et al., 1987; Korolev et al., 
1998). With this in mind we will approximate the length of single stranded chains, 
$Na$ as identical to that of the double stranded ones, $N2b$, denoting both by $L$.
For $N=30$ we thus have $L \approx 100 \mathring{A}$.

The probes are chemically grafted to the surface via a short spacer chain. 
The attainable values of the area per probe, $\Sigma$, vary with the support 
surface (Graves, 1999; Southern et al., 1999; Pirrung, 2002). Typical values of 
$\Sigma$ on glass surfaces are of order of $10^{4} \mathring{A}^{2}$ 
corresponding to a distance $\Delta \approx 
100 \mathring{A}$ between grafting sites. Significantly higher grafting densities 
of ssDNA are possible on polypropylene supports where $\Sigma$ values of $\Sigma
\approx 40 \mathring{A}^{2}$, corresponding to $\Delta \approx 7 \mathring{A}$, 
were reported. In this last case it is necessary to deplete the surface in order 
to allow full hybridization to take place. The mode of grafting can influence the
orientation of the probe. Their orientation can also be affected by adsorption
to the surface (Levicky et al., 1998). Thus ssDNA grafted onto untreated gold
form a compact layer due to adsorption. The layer swells and extends into the
solution following treatment with mercaptohexanol (Levicky et al., 1998). 
This treatment is also important for elimination of non specific adsorption of 
the targets. Our discussion assumes flexible junctions that enable free rotation 
and a non-adsorbing surface. Under this conditions, the average thickness of the 
probe layer, $H$, varies between $H \approx L/2$ at low grafting densities and
$H \approx L$ when $\Sigma \ll L^{2}$.

Three electrostatic length scales are of importance to our discussion. One is
the Bjerrum length $l_{B}=e^{2}/\epsilon kT$ where $\epsilon$ is the dielectric 
constant, $k$ is the Boltzman constant and $T$ is the temperature. In water, 
with $\epsilon \approx 80$, at room temperature, $l_{B} \approx 7 \mathring{A}$. 
Note that the variation of $\epsilon$ with $T$ contributes to the $T$ dependence 
of $l_{B}$. The second is the Gouy-Chapman length $\Lambda = 1/2 \pi l_{B}\sigma$. 
Here $\sigma$ is the number of charges per unit area on a uniformly charged
surface. $\Lambda$ characterizes the spatial distribution of the counterions
in the vicinity of a uniformly charged planar surface in a salt free solvent.
In this situation the majority of counterions are localized within a distance
$\Lambda$ from the surface. In the following the charge of the probes,
hybridized or not, is assumed to be uniformly smeared. As a result, $\sigma$ 
varies between $N/\Sigma$ to $2N/\Sigma$ depending on $ x$, the degree of 
hybridization. For an unhybridized layer $\Lambda$ is in the range of 
$10-10^{2} \mathring{A}$. A third scale is the Debeye length, $r_{D}$, 
characterizing the screening range of electrostatic interactions in a salt 
solution. For a $1:1$ salt with number concentration of ions $\phi_{s}$ 
it is $r_{D} = (8 \pi l_{B} \phi_{s})^{-1/2}$ thus, in a $1M$ solution 
$r_{D}=3 \mathring{A}$.

The range of DNA concentrations encountered in experiments varies in the range
between $10^{-6}M$ to $10^{-12}M$. The solution usually contains also $1M$ of
$1:1$ salt. Under these conditions the electrostatic interactions between the
free targets are essentially fully screened.

\section{The Competition-Free Hybridization Isotherm}

The dependence of the hybridization degree, $x$, on the concentration of the
target, $c_{t}$, is described by the hybridization isotherm. It is helpful to
consider first an array of DNA probes of a single sequence, $p$, in contact
with a solution containing a single species of ssDNA target, $t$. The
hybridization of $p$ and $t$ creates a double stranded oligonucleotide, $pt$,
at the surface. For this choice of system the only reaction is
$p+t \rightleftharpoons pt$ and no competitive hybridization reactions occur
(Figure 1). The factors determining the hybridization isotherm fall into two 
groups. One consists of the factors giving rise to the Langmuir isotherm (Evans 
and Wennerstr\"{o}m, 1994), describing the adsorption of neutral adsorbates at a
surface comprising a finite number of sites, each capable of accommodating a
single adsorbate. These include: (i) the entropy of the free targets in
solution, (ii) the mixing entropy of the hybridized and unhybridized probes
and (iii) the non-electrostatic component of the hybridization free energy.
The hybridization at the surface of a DNA chip differs from the Langmuir
scenario in that both the adsorbates (the targets) and the surface (the probe
layer) are charged. As a result the free energies of the targets and the probe
layer incorporate electrostatic terms. These allow for the electrostatic
interaction energy between the charges and for the entropic effects associated
with the polarization of the ionic clouds surrounding the macroions. 
In the following we will obtain a specific form for the electrostatic free 
energy of the probe layer by modeling it as a planar layer with a laterally 
uniform charge density. However,
some of our conclusions are actually independent of the functional form of
this term. With this in mind we introduce at this point an arbitrary
electrostatic free energy per unit area $\gamma_{el}$. The electric charge
localized at the surface increases with the fraction of hybridized probes,
$x$. Consequently, $\gamma_{el}=\gamma_{el}(x)$ increases with $x$, reflecting
the growth of the electrostatic penalty with the hybridization degree.
Initially we will obtain the hybridization isotherm in terms of this
unspecified $\gamma_{el}(x)$. We will then consider the hybridization
isotherms as obtained for two models for the charge distribution within the
probe layer and the resulting explicit functional forms of $\gamma_{el}(x)$.

The equilibrium state of the hybridization reaction, $p+t \rightleftharpoons
pt$, is determined by the condition $\mu_{pt}=\mu_{p}+\mu_{t}$ where $\mu_{i}$
is the chemical potential of species $i$. Our discussion focuses on the case
where the number concentration of the targets is only weakly diminished by this
reaction and is well approximated by the initial concentration $c_{t}$. Since 
the target solution is dilute and the ionic strength of the solution is high,
electrostatic interactions between the targets are screened. Consequently
$\mu_{t}$ assumes the weak solution form
\begin{equation}
\mu_{t}=\mu_{t}^{0}+kT\ln c_{t} \label{II1}
\end{equation}
where $\mu_{t}^{0}$ is the chemical potential of the reference state. Strictly
speaking, $\mu_{t} = \mu_{t}^{0} + kT \ln a_{t}$ where $a_{t}$ is the activity
(Moore, 1972). The dimensionless $a_{t}$ is related to the concentration
of $t$ chain $c_{t}$ via $a_{t}=\gamma c_{t}$ where $\gamma$ is the activity
coefficient. Since $\gamma \rightarrow 1$ as $c_{t} \rightarrow 0$ we will, for
simplicity express $\mu_{t}$ by Eq.~\ref{II1} noting that $c_{t}$ in this
expression is dimensionless. When the concentration of targets is significantly 
modified by the hybridization with the probes, $c_t$ should be replaced by
$c'_t =  c_t - x N_T/V$ where $V$ is the volume of the solution and $N_T$ the 
total number of probes. Such modification is necessary when $c_t$ is very low 
or when the spots are large.

In order to obtain $\mu_{pt}$ we need to first specify the free energy of the
probe layer as a function of $x$. The $N_{T}$ probes are immobilized at the
surface thus forming a two dimensional grid of hybridization sites. At
equilibrium $N_{pt}=xN_{T}$ of the probes are hybridized while $N_{p}
=(1-x)N_{T}$ remain unhybridized. The $pt$ and $p$ chains form thus a two
dimensional solution associated with a mixing entropy of $-kN_{T}[x\ln
x+(1-x)\ln(1-x)]$. This two dimensional solution is however non-ideal because
of the electrostatic interactions between the chains. Altogether, the free
energy per probe site is
\begin{equation}
\gamma_{site}=\gamma_{0}+x\mu_{pt}^{0}+(1-x)\mu_{p}^{0}+\Sigma \gamma_{el}
+kT[x\ln x+(1-x)\ln(1-x)] \label{Ii2}
\end{equation}
where $\Sigma$ is the area per probe and $\gamma_{0}$ is the free energy 
density of the bare surface. $\mu_{pt}^{0}$ and $\mu_{p}^{0}$ 
are the chemical potentials of the $p$ and $pt$ states in a reference state to 
be discussed later. For simplicity we now limit the discussion to probes and 
targets with identical number of bases, $N$. Since each chain carries a charge 
of $-Ne$, the number charge density on a surface of total area $A$ is 
$\sigma=N(N_{p}+2N_{pt})/A=\sigma_{0}(1+x)$ where $\sigma_{0}=NN_{T}/A$ is 
the number charge density on the unhybridized surface and $\Sigma=A/N_{T}$. 

It is convenient to reformulate the equilibrium condition $\mu_{pt}=\mu_{p}
+\mu_{t}$ in terms of the exchange chemical potential of the hybridized
probe $\mu_{pt}^{ex}=\mu_{pt}-\mu_{p}$. The exchange chemical potential of 
the hybridized probe is $\mu_{pt}^{ex}=\partial \gamma_{site}/\partial x$ or
\begin{equation}
\mu_{pt}^{ex}=\mu_{pt}^{0}-\mu_{p}^{0}+N\frac{\partial \gamma_{el}}
{\partial\sigma}+kT\ln\frac{x}{1-x} \label{II4}
\end{equation}
where $\Sigma\frac{\partial \gamma_{el}}{\partial x}=\Sigma \frac{\partial
\gamma_{el}}{\partial \sigma} \frac{\partial \sigma}{\partial x}=N\frac
{\partial \gamma_{el}}{\partial \sigma}$ since $\partial \sigma/\partial
x=\sigma_{0}$ and $\Sigma \sigma_{0}=N$. $N\frac{\partial \gamma_{el}}
{\partial \sigma}$ is thus the electrostatic free energy penalty incurred 
upon hybridization for a given $x$. The equilibrium condition $\mu_{pt}^{ex}
=\mu_{t}$ then leads to the adsorption isotherm
\begin{equation}
\frac{x}{c_{t}(1-x)}=K_{t}\exp\left[-\frac{N}{kT}\frac{\partial \gamma_{el}}
{\partial \sigma}\right]  \label{II5}
\end{equation}
where $K_{t} = \exp \left( -\frac{\Delta G^{0}}{kT} \right)$ is the 
equilibrium constant for the hybridization reaction at the surface 
and $\Delta G^{0} = \mu_{pt}^{0} - \mu_{p}^{0}-\mu_{t}^{0}$.

Our discussion up to this point did not involve a particular model for the
charge distribution or a specific functional form of $\gamma_{el}$. In the remainder
of this section we will consider the hybridization isotherm for particular forms
of $\gamma_{el}$ as obtained by assuming that the charges of the $p$ and $pt$
chains are uniformly smeared laterally. We will consider two models of this
type. In the first the charges are distributed in an infinitely thin layer at
the solid-liquid interface. This model ignores the structure of the probe
layer and overestimates $\gamma_{el}$. It is however of interest as a simple
model that captures the essential physics. The exact form of $\gamma_{el}$
corresponding to this scenario, for the high salt regime encountered
experimentally, is specified by the Poisson-Boltzman (PB) equation for
$r_{D} \ll \Lambda$ (Evans and Wennerstr\"{o}m, 1994). This $\gamma_{el}$ is
identical to the one obtained by the use of the capacitor
approximation. In this approximation $\gamma_{el}$ is identified with the
electrostatic energy of a planar capacitor, $2\pi(\sigma e)^{2}d/\epsilon$,
with a charge density $\sigma=\sigma_{0}(1+x)$ and a width $d=r_{D}$ thus
leading to
\begin{equation}
\frac{\gamma_{el}}{kT}=2\pi\sigma^{2}l_{B}r_{D}. \label{II6}
\end{equation}
For this choice of $\gamma_{el}$ the hybridization isotherm Eq.~\ref{II5}
assumes the form
\begin{equation}
\frac{x}{c_{t}(1-x)}=K_{t}\exp\left[-\Gamma_{c}(1+x)\right]  \label{II7}
\end{equation}
where $\frac{N}{kT}\frac{\partial \gamma_{el}}{\partial \sigma}=\Gamma_{c}(1+x)$
and $\Gamma_{c}=4\pi N\sigma_{0}l_{B}r_{D}$ is the electrostatic free energy
of a hybridized target in an unhybridized layer with a charge density
$\sigma_{0}$.

The capacitor model accounts for the essential physics in a simple and
transparent way. However this model tends to overestimate the electrostatic
free energy because all the charges of the DNA chains are placed on a surface.
To avoid this problem we now assume instead that the charges are uniformly
smeared within a layer of thickness $H$ giving rise to a number charge density
of $\rho=\sigma/H$. The analysis of this ``diffuse layer'' model differs from 
that of the capacitor model only in the form of the electrostatic free energy 
density $\gamma_{el}$. To obtain $\gamma_{el}$ we utilize a two phase or a 
``box'' approximation for the solution of the PB equation (Pincus, 1991; 
Wittmer and Joanny, 1993; Borisov et al., 1994). Within it,
we distinguish between two regions: (i) a proximal region, adjacent to the
charged surface, where the concentrations of ions deviates from the bulk
values. The concentrations of each of the ionic species are constant and obey
the Donnan equilibrium. (ii) A distal neutral region where the effect of the
charged surface is screened out and the concentrations of the ions are
determined by the concentration of the salt. The ionic concentrations and the
equilibrium electrostatic free energy are determined by minimization of the
free energy with respect to the height of the proximal region. This
approximation involves the simplest form of discretization of the PB equation.
The details of the analysis are presented in the Appendix A. In the following
we focus on the experimentally relevant case of high salt such that 
$r_{D} \ll H$ and $r_{D} \ll (H\Lambda)^{1/2}$. The low salt regime is 
described in Appendix B. In the high salt regime the screening of the 
charged layer is dominated by the contribution of the salt and
\begin{equation}
\frac{\gamma_{el}}{kT}=4 \pi \sigma^{2}l_{B}\frac{r_{D}^{2}}{H}. \label{III1}
\end{equation}
The hybridization isotherm in this ``salt screening'' (ss) regime is
\begin{equation}
\frac{x}{c_{t}(1-x)}=K_{t}\exp\left[-\Gamma(1+x)\right]  \label{III2}
\end{equation}
where $\frac{N}{kT}\frac{\partial \gamma_{el}}{\partial \sigma} \approx  8 \pi N
\sigma l_{B}\frac{r_{D}^{2}}{H}=\Gamma(1+x)$ and $\Gamma=8\pi N\sigma_{0}l_{B}
\frac{r_{D}^{2}}{H}$ is the electrostatic penalty incurred by a $pt$ chain in
an unhybridized layer with $\sigma=\sigma_{0}$. Note that the functional form
of Eq.~\ref{III2} is identical to that of Eq.~\ref{II7} but $\Gamma=2\Gamma_{c}
r_{D}/H<\Gamma_{c}$.

As a reference state it is convenient to choose the state of a chain (ssDNA or
dsDNA) anchored to a surface at a low grafting density such that the in-plane
electrostatic interaction are negligible. When the ``lateral''
interactions are negligible, one may roughly approximate $\mu_{pt}^{0}$
($\mu_{p}^{0}$) by the $\mu^{0}$ of the corresponding free chain in the
solution. This choice is useful in that it enables us to estimate the various
hybridization constants using the nearest neighbor parameter sets available in
the literature (Bloomfield et al., 2000). It is however important to keep in
mind the problems introduced by this choice of reference state and the
approximation of $\mu_{pt}^{0}$ ($\mu_{p}^{0}$). One difficulty involves the
electrostatic free energy. $\gamma_{el}$ is obtained by charging of a
hypothetical non charged layer. As a result, the electrostatic contribution to
$\mu_{pt}^{0}$ ($\mu_{p}^{0}$) leads to a small overestimate of the
electrostatic free energy. Note that for high $\sigma$ or small $\Lambda$ 
fluctuation effects become important (Lau et al., 2002). These are not included
in our analysis. In addition, caution is required in
identifying the boundaries of the regime of negligible lateral interactions.
This is because the decay of electrostatic interactions at an impenetrable
surface is slower than in the bulk. Thus, point charges embedded at an
impenetrable surface polarize an hemisphere of the ionic solution thus giving
rise to a dipole and the lateral interactions decay as $1/r^{3}$ (Jancovici,
1982). Another problem concerns the rotational free energy of the chains.
The rotational freedom of the terminally anchored chains is restricted by the
impenetrable grafting surface. Further restrictions may be imposed by the
grafting functionality. The diminished rotational freedom reduces the
rotational term in the free energy per chain. This effect is however neglected
when $\mu_{pt}^{0}$ ($\mu_{p}^{0}$) are approximated by $\mu^{0}$ of the
corresponding free chains. When both the target and probe are self complementary 
it is necessary to allow for the change of symmetry due to the grafting. In turn,
this requires an appropriate modification of $\mu_{pt}^{0}$ ($\mu_{p}^{0}$)
with respect to their bulk counterparts. Finally, note that in the low
grafting density regime, as discussed above, the hybridization isotherm is
expected to assume the Langmuir form
\begin{equation}
\frac{x}{(1-x)c_{t}}=K_{t}. \label{III5i}
\end{equation}
In this regime the electrostatic aspect of the problem is evident only in the
dependence of the $\mu^{0}$s and thus $K_{t}$, on the concentration of salt.

\begin{figure}
\centerline{\epsfig{file=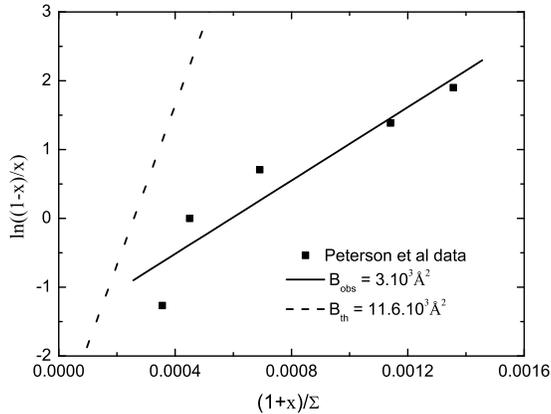,width=3.25in}}
\caption{A plot of $\ln(1-x)/x$ vs. $(1+x)/\Sigma$ using the data of Peterson
et al. (2002). Eq.~\ref{III2} yields $\ln(1-x)/x=const^{\prime}+B(1+x)/\Sigma$
with $B=8\pi l_{B}N^{2}r_{D}^{2}/H$. For the experiment cited $l_{B}=7
\mathring{A}$, $r_{D}=3\mathring{A}$, $N=25$ and $H=85 \mathring{A}$ leading
to $B \simeq 1.16 \times 10^{4}\mathring{A}^{2}$ as compared to the observed
$B \simeq 3 \times 10^{3}\mathring{A}^{2}$.}
\end{figure}

The number of hybridization isotherms of DNA chips reported in the literature
is rather small (Nelson et al., 2001; Peterson et al., 2001; Peterson et al., 
2002). The situation is further complicated because of paucity of data concerning
$N_{T}$, the number of probes available to hybridization, and the related
problem of ascertaining the attainment of thermodynamic equilibrium. The
``uniform smearing'' models for the hybridization isotherms are
supported by two experimental studies carried out by the group of Georgiadis.
In one experiment the grafting density was varied in the range of
$2 \cdot 10^{12} - 12 \cdot 10^{12}$ probes/$cm^{2}$ while $c_{t}$ was kept 
constant at $1 \mu M$ (Peterson et al., 2001). A plot of $\ln[(1-x)/x]$ vs. 
$(1+x)/\Sigma$ can be fitted with a straight line with a slope smaller than the 
one predicted by the theory (Figure 4). This is however encouraging since the 
data was acquired  in $1/2h$ and is thus unlikely to reflect complete equilibrium. 
In the second group of experiments, the hybridization was studied for a lower
grafting density of $1.5 \cdot 10^{12}$ probes/$cm^{2}$ while $c_{t}$ was varied
over the range of $500mM$ to $5\mu M$ (Peterson et al., 2002). In this study the
hybridization isotherm of the perfectly matched targets was well fitted by
the Langmuir form. Importantly, this study established that the system failed 
to reach equilibrium without heating treatment for mismatched targets.

A hybridization isotherm of identical form to Eq.~\ref{II7} and to Eq.~\ref{III2}
was announced earlier by Vainrub and Pettitt (Vanirub and Pettitt, 2002;
Vanirub and Pettitt, 2003). VP also pointed out that some of the results of the
Georgiadis group are consistent with this form. The VP approach is designed to
permit the utilization of exact results on the interaction free energy between
a penetrable charged sphere and an impenetrable charged surface in the strong
screening regime when the Debye-H\"uckel approximation is applicable (Ohshima and 
Kondo; 1993). Within it, one calculates the excess free energy of a probe layer 
with $x N_T$ hybridized probes, $F_{el}(x)$, with respect to the unhybridized 
layer. In effect, $F_{el}(x)$ is the sum of the contributions of $x N_T$ 
hybridization events, $F_{el} = \sum_{i=1}^{x N_T} F_i (\sigma_i)$. 
Each step contributes $F_i (\sigma_i) = F_{pt}(\sigma_i) - F_p(\sigma_i)$ 
where $F_{pt}(F_p)$ is the electrostatic free energy of a $pt$ $(p)$ sphere 
in contact with a planar layer with charge density $\sigma_i = \sigma_0 + i N/A$. 
Thus, at each step the probe layer is modeled as a planar charged surface 
interacting  with a {\it single} charged sphere. The steps differ in the 
charge density of the surface. The main difference between the VP approach 
and ours is in the handeling of the charges. In the VP scheme some of the 
charges appear as charged spheres while others appear as a charged surface. 
Within our model 
there is no duality and all charges are described in the same fashion. 
In practical terms, the VP approach can not allow for the thickness of the 
probe layer nor can it be extended to describe hybridization at lower ionic 
strength.

\section{Sensitivity, Selectivity and $c_{50}$ for Competition Free Systems}

The hybridization isotherms discussed in the two preceding sections describe
DNA arrays in the absence of competitive hybridization in the bulk or at the
surface. This situation is realized when an array comprising of single type of
probes is exposed to a solution of a single target. The concentration of
target leading to $50\%$ equilibrium hybridization in such systems,
$^{t}c_{50}^{0}=K_{t}^{-1}\exp(\frac{N}{kT}\frac{\partial \gamma_{el}}
{\partial \sigma}|_{x=1/2})$ is a useful characteristic of the system. Within
the diffuse layer model in the salt screening (ss) regime $^{t}c_{50}^{0}$ is
\begin{equation}
^{t}c_{50}^{0}=\frac{1}{K_{t}}\exp(\frac{3}{2}\Gamma). \label{IIIa1}
\end{equation}
$^{t}c_{50}^{0}$ is closely related to the sensitivity of the array
\begin{equation}
S_{e}(x)=\frac{(1-x)^{2}}{1+x(1-x)\Gamma}K_{t}\exp[-\Gamma(1+x)]=\frac
{(1-x)^{2}}{1+x(1-x)\Gamma}\frac{1}{^{t}c_{50}^{0}}\exp[-\Gamma(x-1/2)].
\label{IIIa2}
\end{equation}
The sensitivity of the array, as defined by $S_{e}(x)$ varies with $x$ and
thus with $c_{t}$. It is maximal at $x=0$ when
\begin{equation}
S_{e}(0)=K_{t}\exp(-\Gamma)=\frac{1}{^{t}c_{50}^{0}}\exp(-\frac{\Gamma}{2})
\label{IIIa3}%
\end{equation}
while at $x=1/2$ it is $S_{e}(1/2)=1/(4+\Gamma)^{t}c_{50}^{0}$. As we shall see,
$S_{e}(0)$ is not affected by competitive hybridization. On the other hand,
$S_{e}(x)$ and $c_{50}$ are modified significantly by these processes.

Since $S_{e}(x) \sim 1/^{t}c_{50}^{0}$, clearly a lower $^{t}c_{50}^{0}$ is 
desirable and $1/^{t}c_{50}^{0}$ is a useful measure of the sensitivity 
of the array. Both $1/^{t}c_{50}^{0}$ and $S_{e}(x)$ decrease as
$\Gamma$ and the electrostatic penalty incurred by the hybridization increase. 
In the salt screening regime, where most experiments are carried
out, $\Gamma$ increases with the grafting density as $\Gamma \sim \sigma_{0}$.
While higher sensitivity is expected at lower grafting densities, this does
not ensure a lower detection limit or a better quantitative resolution. These
last two parameters depend also on the measurement error $\epsilon_{r}$. 
In turn, $\epsilon_{r}$ typically decreases as the grafting density, and the 
signal, increase. Thus, $1/^{t}c_{50}^{0}$ and $S_{e}(x)$ only provide partial
guidance for the design of DNA arrays. Nevertheless, these two parameters do
provide useful information regarding the performance of a DNA chip of a given
design (that is, grafting density, grafting functionality, spot size and
detection method). Thus, the relative sensitivity of two different probe
target pairs, $p_{1}$ $t_{1}$ and $p_{2}$ $t_{2}$, all other factors being
equal, is
\begin{equation}
\frac{^{t_{1}}S_{e}}{^{t_{2}}S_{e}}=\frac{^{t_{2}}c_{50}^{0}}{^{t_{1}}
c_{50}^{0}}=\frac{K_{t_{1}}}{K_{t_{2}}}. \label{IIIa4}
\end{equation}
The specificity of a given probe, $p$, can be quantified by the relative
sensitivity when a $p$ spot is exposed to a perfectly matched target, $t$, or
to a mismatch, $m$
\begin{equation}
\frac{^{t}S_{e}}{^{m}S_{e}}=\frac{^{m}c_{50}^{0}}{^{t}c_{50}^{0}}=\frac{K_{t}
}{K_{m}}. \label{IIIa5}
\end{equation}
These two ratios also specify the corresponding ratios of the qualitative
resolution and the detection limit. Importantly, Eq.~\ref{IIIa4} and
Eq.~\ref{IIIa5} are independent of the electrostatic penalty irrespective of the
form of $\gamma_{el}$.

\section{The Effect of Competitive Surface Hybridization}

The hybridization isotherm requires modification when the bulk solution
contains more than one ssDNA species capable of hybridization at the surface.
In this situation the different species compete for hybridization with the
probes. For simplicity we consider the case of a binary solution
comprising a target ($t$) and a mismatched ssDNA ($m$) with a concentration
$c_{m}$ and a standard chemical potential in the bulk solution $\mu_{m}^{0}$.
It is placed in contact with a single component probe layer such that the $p$
chains are perfect matches to the targets (Figure 2). We further assume that 
the $m$ and $t$ chains are of the same length. The number of probes that 
hybridized with $m$ is $N_{m}=yN_{T}$. In this case $\sigma = N 
(N_{p}+2N_{pt}+2N_{pm})/A = \sigma_{0}(1+x+y)$ and
\begin{equation}
\gamma_{site}=\gamma_{0}+x\mu_{pt}^{0}+y\mu_{pm}^{0}+(1-x-y)\mu_{p}^{0}
+\Sigma\gamma_{el}+kT[x\ln x+y\ln y+(1-x-y)\ln(1-x-y)] \label{IV1}
\end{equation}
where $\mu_{pm}^{0}$ is the standard chemical potential of a hybridized $pm$
at the surface. The hybridization isotherm in this situation is determined by 
two equilibrium conditions $\mu_{pt}^{ex}=\mu_{t}$, as before, and $\mu_{pm}^{ex}
=\mu_{m}$. In obtaining the explicit form of these conditions note that 
$\frac{\partial \Sigma \gamma_{el}} {\partial x}=\frac{\partial \Sigma 
\gamma_{el}}{\partial y}=N\frac{\partial \gamma_{el}}{\partial \sigma}$ 
because $\frac{\partial \sigma}{\partial x}=\frac{\partial \sigma}{\partial y}=
\sigma_{0}$. The exchange chemical potentials of the hybridized $m$ and $t$ are 
thus given by
\begin{equation}
\mu_{pt}^{ex}=\mu_{pt}^{0}-\mu_{p}^{0}+N\frac{\partial\gamma_{el}}
{\partial\sigma}+kT\ln\frac{x}{1-x-y} \label{IV2}
\end{equation}%
\begin{equation}
\mu_{pm}^{ex}=\mu_{pm}^{0}-\mu_{p}^{0}+N\frac{\partial\gamma_{el}}
{\partial\sigma}+kT\ln\frac{y}{1-x-y} \label{IV3}
\end{equation}
and the chemical potential of the free $m$ is
\begin{equation}
\mu_{m}=\mu_{m}^{0}+kT\ln c_{m}. \label{IV4}
\end{equation}
As before, we focus on the ``small spot'' limit where the bulk concentrations 
of $m$ and $t$ are not affected by the hybridization at the surface.
The hybridization behavior of this system is described by three isotherms
specifying the hybridization degrees of $t$ and $m$ individually as well as
the total hybridization:
\begin{equation}
\frac{x}{c_{t}(1-x-y)}=K_{t}\exp\left[-\frac{N}{kT}\frac{\partial
\gamma_{el}}{\partial \sigma}\right]  \label{IV5}
\end{equation}
\begin{equation}
\frac{y}{c_{m}(1-x-y)}=K_{m}\exp\left[-\frac{N}{kT}\frac{\partial
\gamma_{el}}{\partial \sigma}\right]  \label{IV6}
\end{equation}
\begin{equation}
\frac{x+y}{(1-x-y)}=\left(c_{m}K_{m}+c_{t}K_{t}\right) \exp\left[-\frac{N}{kT}
\frac{\partial \gamma_{el}}{\partial \sigma}\right]  \label{IV7}
\end{equation}
where $K_{t}=\exp\left(-\frac{\Delta G^{0}}{kT}\right)$, $K_{m}=
\exp\left(-\frac{\Delta G_{m}^{0}}{kT}\right)$ and $\Delta G_{m}^{0}
=\mu_{pm}^{0}-\mu_{p}^{0}-\mu_{m}^{0}$. The observed isotherm depends on 
the method used to interrogate the surface. Thus, utilization of 
selectively tagged $t$ will reveal Eq.~\ref{IV5}, use of selectively 
tagged $m$ will show Eq.~\ref{IV6} while detection
methods sensitive to overall hybridization mass, such as surface plasmon
resonance, will yield Eq.~\ref{IV7}. The explicit form of the
hybridization isotherms within the diffuse model in the salt screening regime
is obtained by substituting $\frac{N}{kT}\frac{\partial \gamma_{el}}
{\partial \sigma}=\Gamma(1+x+y)$. Note that $K_{t}$, $K_{m}$ and $\Gamma$ can 
be determined from experiments involving exposure of the DNA chip to single 
component solutions of $t$ and $m$ chains.

\begin{figure}
\centerline{\epsfig{file=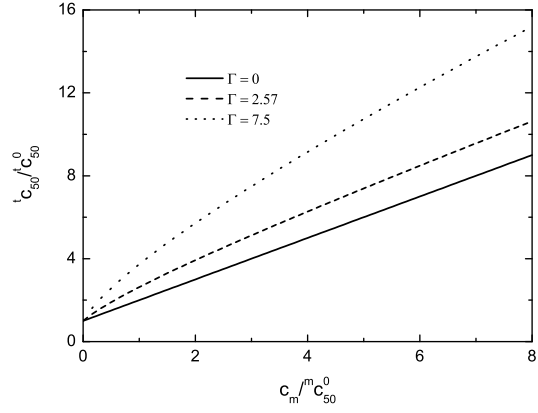,width=3.25in}}
\caption{Plots of $^{t}c_{50}/^{t}c_{50}^{0}$ vs. $c_{m}/^{m}c_{50}^{0}$, as
given by Eq.~\ref{IV9}, for the case of competitive surface hybridization
involving a probe, $p$, of the sequence CAACTTGATATTAATA, a target, $t$,
GTTGAACTATAATTAT and a mismatched target, $m$, GTTGA\underline{G}CTATAATTAT (TG mismatch).
In the three cases depicted $T=300^{\circ}K$, $N=16$, $H=54 \mathring{A}$,
$l_{B}=7 \mathring{A}$ and $r_{D}=3 \mathring{A}$. The continuous line
corresponds to the low grafting density regime where $\Gamma=0$. The two other
are $\Sigma=H^{2}=2916 \mathring{A}^{2}$ leading to $\Gamma=2.57$ (dashes) and
$\Sigma = 10^{3} \mathring{A}^{2}$ leading to $\Gamma=7.5$ (dots). The standard
Gibbs free energies per mole at $37^{\circ}C$ are $\Delta G_{t}^{0}=12.4kcal/mole$
and $\Delta G_{m}^{0}=10.1kcal/mole$ (Tibanyenda et al., 1984). Since the
$\Delta G^{0}$ are per mole rather than per molecule, the equilibrium
constants at $T=300^{\circ}K,$ neglecting the $T$ dependence of the $\Delta G^{0}$,
are $K_{t}=\exp(-\Delta G_{t}^{0}/RT) \simeq 10^{9.0}$ and $K_{m}=\exp(-\Delta
G_{m}^{0}/RT) \simeq 10^{7.4}$ where $R$ is the gas constant. The corresponding
$^{t}c_{50}^{0}$ are $10^{-9}M, 10^{-7.4}M$ and $10^{-4.1}M$ respectively. 
The values of $^{m}c_{50}^{0}$ are $10^{-7.4}M, 10^{-5.7}M$ and $10^{-2.5}M$}
\end{figure}

The specificity of the assay can be quantified by considering the fraction of
``incorrectly'' hybridized probes, $P_{m}$. Equations \ref{IV5} and \ref{IV6} 
yield $y=x\frac{c_{m}}{c_{t}}\frac{K_{m}}{K_{t}}$ and thus
\begin{equation}
P_{m}=\frac{y}{x+y}=\frac{c_{m}K_{m}}{c_{m}K_{m}+c_{t}K_{t}}. \label{IV8}
\end{equation}
Within this definition the specificity strongly depends on $c_{m}$, or to be
precise on $\frac{c_{m}}{c_{t}}\frac{K_m}{K_{t}}$. The fraction of mismatched
probes is small, $P_{m} \ll 1$, so long as $c_{m}\ll c_{t}\frac{K_m}{K_{t}}$. 
At $c_{m}=c_{t}\frac{K_m}{K_{t}}$, half of the hybridized probes are mismatched,
$P_{m}=1/2$, while for $c_{m} \gg c_{t}\frac{K_m}{K_{t}}$, $P_{m}$ approaches
unity. Equation \ref{IV8} is independent of the electrostatic contribution
irrespective of the form of $\gamma_{el}$. It is also useful to consider the
ratio of $^{t}c_{50}^{0}$ to $^{t}c_{50}$, the bulk concentration of $t$ giving 
rise to $50\%$ $pt$ hybridization in the presence of a mismatch of concentration 
$c_{m}$. In contrast to $P_{m}$, the expression for $^{t}c_{50}^{0}/^{t}c_{50}$ 
does depend on $\gamma_{el}$. For the diffuse layer model in the salt screening 
regime it is given by
\begin{equation}
\frac{^{t}c_{50}^{0}}{^{t}c_{50}}=\left(1-\frac{c_{m}}{^{m}c_{50}^{0}}
\frac{^{t}c_{50}^{0}}{^{t}c_{50}}\right)  \exp\left(-\frac{\Gamma}{2}
\frac{c_{m}}{^{m}c_{50}^{0}}\frac{^{t}c_{50}^{0}}{^{t}c_{50}}\right).
\label{IV9}
\end{equation}
In the low grafting density regime, when $\Gamma=0$, Eq.~\ref{IV9} assumes the
form $\frac{^{t}c_{50}}{^{t}c_{50}^{0}}=1+\frac{c_{m}}{^{m}c_{50}^{0}}$. In
all cases, $^{t}c_{50}=  {}^{t}c_{50}^{0}$ when $c_{m}=0$ and $^{t}c_{50}>
{}^{t}c_{50}^{0}$ for $c_{m}>0$. In other words, the sensitivity, as
measured by $1/^{t}c_{50}$, decreases as $c_{m}$ increases (Figure 5).

\section{The Effect of Competitive Bulk Hybridization}

A different type of competition occurs when the targets can hybridize in the
bulk as well as at the surface. Such competition can arise in three different
situations: (i) The solution contains targets as well as complementary
strands, $c$. These can be perfectly matched or mismatched sequences. The $c$
chains hybridize with the targets to form free double stranded $tc$ DNA chains.
Thus, the $t+c \rightleftharpoons tc$ reaction in the bulk competes with
the $t+p \rightleftharpoons pt$ reaction at the surface (Figure 3). 
(ii) The targets are self-complementary and thus capable of undergoing a bulk 
hybridization reaction $t+t \rightleftharpoons tt$ in addition to 
$t+p \rightleftharpoons pt$ where $p$ now denotes the immobilized $t$ probe. 
(iii) A third possible scenario involves formation of hairpins.

As explained in section III, within our discussion the lengths of the $p$ and $pt$ 
chains are identical. Accordingly we will focus on the first two cases where
the length of the chains does not change upon hybridization. Initially, we
discuss the $t+c \rightleftharpoons tc$ scenario and then comment on the
modification required to adapt the analysis to the $t+t \rightleftharpoons tt$
case. Again, we focus on the ``small spot'' limit assuming that the hybridization 
with the probes has a negligible effect on the concentration of the targets. 
The hybridization isotherm describing this situation, for the two cases of 
interest, is
\begin{equation}
\frac{x}{(1-x)[t]}=K_{t}\exp\left[-\frac{N}{kT}\frac{\partial\gamma_{el}}
{\partial\sigma}\right]  \label{V1}
\end{equation}
and $\frac{N}{kT}\frac{\partial \gamma_{el}}{\partial \sigma}=\Gamma(1+x)$ 
in the ss regime of the diffuse layer model. Importantly, the hybridization
isotherm is modified in that $c_{t}$, the total concentration of $t$, is
replaced by the equilibrium $t$ concentration, $[t]$. In turn, $[t]$ is
determined by the mass action law governing the bulk hybridization reaction.
The combination of Eq.~\ref{V1} with the appropriate mass action law is
equivalent to the equilibrium condition specified by $\mu_{t}+\mu_{p}=
\mu_{pt}$ and $\mu_{t}+\mu_{c}=\mu_{tc}$.

In the $t+c \rightleftharpoons tc$ scenario the mass action law is
$[tc]/[t][c]=K$ where $[i]$ is the equilibrium concentration of species
$i$ and $K$ is the equilibrium constant of the bulk hybridization 
reaction for the temperature and ionic strength
considered. This is supplemented by the mass conservations relations
$[t]+[tc]=c_{t}$ and $[c]+[tc]=c_{c}$ where $c_{i}$ denotes the total
concentration of $i$. $[t]$ is then specified by
\begin{equation}
K[t]^{2}+\{K(c_{c}-c_{t})+1\}[t]-c_{t}=0. \label{V2}
\end{equation}
When the hybridization with the probes has a significant effect on the 
concentration of the targets, $[t]+[tc] = c_t$ should be replaced by
$[t] + [ct] + x N_T/V = c_t$. For brevity, we will not consider this case.
It is instructive to analyze the effect of the competitive bulk hybridization
for a number of simple situations. When the equilibrium favors the reactants,
$[t]\approx c_{t}$ and the hybridization isotherm retains the competition-free
form, Eq.~\ref{II5}. Such is the case in the presence of large excess of $t$,
$c_{t}\gg c_{c}$ or when $K$ is sufficiently small i.e., $c_{c}\gg c_{t}$ or
$c_{c}\approx c_{t}$ but $K c_{c}\ll 1.$ Significant modification of the
hybridization isotherm occur when the bulk hybridization equilibrium favors
the products. This situation occurs in two simple cases: when $Kc_{c}\gg 1$ 
with either $c_{c}\gg c_{t}$ or $c_{c}\approx c_{t}$. 
We initially discuss briefly the first situation when
\begin{equation}
[t] \approx \frac{c_{t}}{Kc_{c}}\ll c_{t} \label{V3}
\end{equation}
leading to
\begin{equation}
\frac{x}{(1-x)}=\frac{c_{t}}{Kc_{c}}K_{t}\exp\left[-\frac{N}{kT}
\frac{\partial\gamma_{el}}{\partial\sigma}\right].  \label{V4}
\end{equation}
To obtain an explicit form of the isotherm within the ss regime of the diffuse
layer model we substitute $\Gamma (1+x)$ for $\frac{N}{kT}\frac{\partial
\gamma_{el}}{\partial \sigma}$. However, the effect on $^{t}c_{50}$ is
independent of the model. In comparison to $^{t}c_{50}^{0}=K_{t}^{-1}
\exp\left(\frac{N}{kT}\frac{\partial \gamma_{el}}{\partial \sigma}
|_{x=1/2}\right)$, $^{t}c_{50}$ increases to
\begin{equation}
^{t}c_{50} = K c_{c} {}^{t}c_{50}^{0} \gg {}^{t}c_{50}^{0}. \label{V5}
\end{equation}
The sensitivity, as measured by $1/^{t}c_{50}$, is thus reduced by a factor of
$Kc_{c}\gg 1$. When $c_{c}\approx c_{t}$ and $Kc_{c}\gg 1$ the equilibrium
condition (\ref{V2}) yields
\begin{equation}
[t] \approx \left(\frac{c_{t}}{K}\right)^{1/2} \label{V6}
\end{equation}
thus leading to
\begin{equation}
\frac{x}{(1-x)} = \left(\frac{c_{t}}{K} \right)^{1/2} K_{t} \exp \left[
-\frac{N}{kT} \frac{\partial\gamma_{el}}{\partial\sigma}\right].  \label{V7}
\end{equation}
The corresponding $^{t}c_{50}$ increases thus to
\begin{equation}
^{t}c_{50}=K(^{t}c_{50}^{0})^{2} \label{V8}
\end{equation}
and the sensitivity, as measured by $1/^{t}c_{50}$, is reduced by a factor of
$K {}^{t}c_{50}^{0} \gg 1$ in comparison to the competition-free scenario.
The sensitivity $S_{e}=dx/dc_{t}$ does depend on the form of $\gamma_{el}$.
When Eq.~\ref{V8} is applicable $S_{e}$, as specified by the uniform density
model at the ss regime, is
\begin{equation}
S_{e}=\frac{K_{t}^{2}}{4K}\exp[-2\Gamma(1+x)]\frac{(1-x)^{3}}
{x[1+\Gamma x(1-x)]}=\frac{1}{^{t}c_{50}}\exp[-\Gamma(x-1)]\frac{(1-x)^{3}}
{x[1+\Gamma x(1-x)]}. \label{V9}
\end{equation}
However, in the limit of $c_{t} \rightarrow 0$ the effect of the competitive
bulk hybridization is negligible and $S_{e}(0)$ is thus given by
Eq.~\ref{IIIa3}. This is also the case for the $c_{c}\gg c_{t}$ and $K c_{c}
\gg 1$ scenarios considered earlier.

In the low grafting density regime, when $\gamma_{el}$ is independent of
$\sigma$, the hybridization isotherm for $c_{c}\approx c_{t}$ with $Kc_{c}
\gg 1$ assumes the form $x/(1-x)=K_{t} \, (c_{t}/K)^{1/2}$. Upon defining
$K_{eff}=K_{t}^{2}/K$ this isotherm can be expressed as
\begin{equation}
x=\frac{(K_{eff}c_{t})^{1/2}}{1+(K_{eff}c_{t})^{1/2}}. \label{V10}
\end{equation}
This form is of interest because it resembles the isotherm obtained from the
Sips model (Sips, 1948). The Sips model provides a generalization of the
Langmuir isotherm in which the single binding energy, utilized in the Langmuir
version, is replaced by a distribution of binding energies thus leading to an
expression of the form
\begin{equation}
x=\frac{(K_{eff}c_{t})^{a}}{1+(K_{eff}c_{t})^{a}} \label{V11}
\end{equation}
where $a$ is a characteristic of the distribution function. Thus, competitive
bulk hybridization can give rise to a ``Sips isotherm'' with $a=1/2$ even 
though the underlying mechanism is completely different. This is
of interest because the Sips isotherm was recently reported to allow for
improved fitting of hybridization data (Peterson et al., 2002).

When the competitive bulk hybridization involves self-complementary chains,
$t+t \rightleftharpoons tt$, the preceding discussion requires modification. In
this case the mass action law assumes the form $[tt]/[t]^{2}=K$ and the
corresponding mass conservation relation becomes $[t]+2[tt]=c_{t}$. $[t]$ is
thus determined by $2K[t]^{2}+[t]-c_{t}=0$. When $Kc_{t}\ll 1$ the competitive
effect is negligible and $[t]\approx c_{t}$. In the opposite limit, $Kc_{t}
\gg 1$, the bulk hybridization is important and $[t]\approx (c_{t}/2K)^{1/2}$.
The $t+t \rightleftharpoons tt$ scenario thus closely resembles the
$t+c \rightleftharpoons tc$ case when $c_{t}\approx c_{c}$. Note however that
care must be taken in estimating $K_{t}$ for the self complementary case. When
the sequences of the $p$ and $t$ chains are identical, $K_{t}$ differs from the 
bulk $K$ because the grafting to the surface modifies the symmetry of the chain
(in addition to the factors discussed in section IV).

\section{Discussion}

The hybridization isotherms of DNA chips provide a natural starting point for
the analysis of their sensitivity and specificity. Clearly, this description
is incomplete in that it is limited to equilibrium states while in typical
experiments equilibrium is not attained. The hybridization isotherms are
nevertheless of interest because of the emerging evidence that the best
performance of DNA chips is obtained in thermodynamic equilibrium (Bhanot et
al., 2003). Accordingly, the selectivity and specificity obtained from the
hybridization isotherms provide upper bounds to the performance of these
assays. This approach is also of interest because an understanding of the
equilibrium state is a prerequisite for the full analysis of the kinetics of
hybridization. When selectivity is discussed in terms of the slope of the
response curve, it is necessary to use an explicit form of the hybridization
isotherm. We obtained such an explicit expression by use of the diffuse layer
model. In this model the charges of the $pt$ and $p$ chains are uniformly
smeared within the probe layer. However, the analysis of the hybridization
isotherm also suggests the use of various $c_{50}$s as measures of the
specificity and selectivity of DNA chips. This description affords an
important advantage in that the effects of competitive hybridization can be
described in a form that is independent of the model used to specify the
electrostatic interactions. Thus, the best performance of DNA chips is
attained in competition-free situations used to define $^{t}c_{50}^{0}$,
$^{m}c_{50}^{0}$ etc. One can then analyze the effects of competitive
hybridization in terms of the increase in $^{t}c_{50}$ in comparison to
$^{t}c_{50}^{0}$. This analysis also indicates that the knowledge of the
competition-free isotherms allows to predict the isotherms realized when
competitive hybridization occurs. In addition the observed isotherm 
depends on the measurement technique when competitive surface hybridization 
is important i.e., label free detection differs from the detection of 
selectively labeled targets.

Much of our discussion concerns the effects of competitive hybridization. In
certain applications the effects of competitive surface hybridization can be
minimized by proper design of the probes (Lockhart et al., 1996; Li and Stormo, 
2001; Bhanot et al., 2003). Such is the case, for example, when
studying the expression level of genes of known sequence. However, this
strategy can not be employed when DNA chips are used to identify single
nucleotide polymorphism or point mutations. Probe design is also of limited
value in counteracting the effects of competitive bulk hybridization.

The results we obtained are based on the equilibrium hybridization isotherms.
They are formulated in terms of the equilibrium fractions $x$, $y$ etc. of
hybridized probes. In confronting these predictions with experimental results
it is important to note the following two points. First, in order to specify 
$x$ and $y$ it is necessary to determine first the number of probes
\textit{available} to hybridization, $N_{T}$. Thus it is not sufficient to
ascertain the number of $p$ chains immobilized at the surface. It is also
necessary to confirm that this corresponds to the number of hybridized probes
at equilibrium with a large excess of targets. This brings us to the second point
concerning the equilibrium state. This plays a role both in the determination
of $N_{T}$, as discussed above, and in the determination of equilibrium
fractions of hybridized probes. Here we recall again that a stationary state does
not necessarily imply equilibrium. An equilibrium state should also be
independent of the preparation method or sample history. In the context of DNA
chips it is thus important to verify that the stationary state is not affected
by a heating treatment. In every case, the equilibration time can be very long
with periods of up to $14$ hours reported in the literature. It is also useful
to note that the equilibration time depends on the bulk composition, $c_{t}$
and $c_{m}$, on the ionic strength and the grafting density, $\Sigma$. It also
varies with the number of mismatches and their identity. Accordingly, the
equilibration time in one experimental situation is not necessarily identical
to the equilibration time under different conditions. When studying
simultaneously the hybridization on different spots the equilibration rates
for the different spots may well differ.

It is useful to distinguish between two types of experiments involving DNA 
chips: experiments designed to elucidate the physical chemistry of their 
function and experiments utilizing DNA chips to analyse biological samples. 
In the first category, the experimental set up allows for selective 
labeling and for precise control of the composition of the bulk solution.
It is straightforward to confront our analysis with such ``physical 
chemistry'' experiments. The situation with respect to analytical 
applications is more complex. Analytical experiments typically rely on PCR
amplification of biological samples. As a result, selective labeling is 
impossible and the composition of the bulk solution is determined by
the composition of the original sample and the amplification scheme i.e., 
the choice of primers. Our discussion reveals difficulties in the quantitative 
interpretation of the results of such experiments, especially when 
used to study point mutations. In this last situation, one may quantify
errors introduced by the competitve hybridization by use of ``standard 
addition'' i.e., study a series of solutions obtained from the amplified
biological sample by addition of different amounts of synthetic, 
selectively labeled target. The practical importance of these difficulties
and the methods to overcome them remain to be established. 

\section{Appendix A: The Box Model for a Diffuse and for a Planar Layer.}

We consider a diffuse layer carrying $Q$ charges distributed uniformly in a
region of height $H$ such that the total charge is $-Qe<0$. The resulting
number charge density is $\rho=Q/AH=\sigma/H$ where $\sigma=Q/A$ is the
corresponding surface number density of charges and $A$ is total
surface area. In the limit of $H=0$ this system reduces to the case of a
charged surface. The analytical solution of the PB equation for this last case
is known. Accordingly we will also investigate the box model for the $H=0$ in
order to demonstrate that it recovers the known results up to a numerical factor.

\begin{figure}
\centerline{\epsfig{file=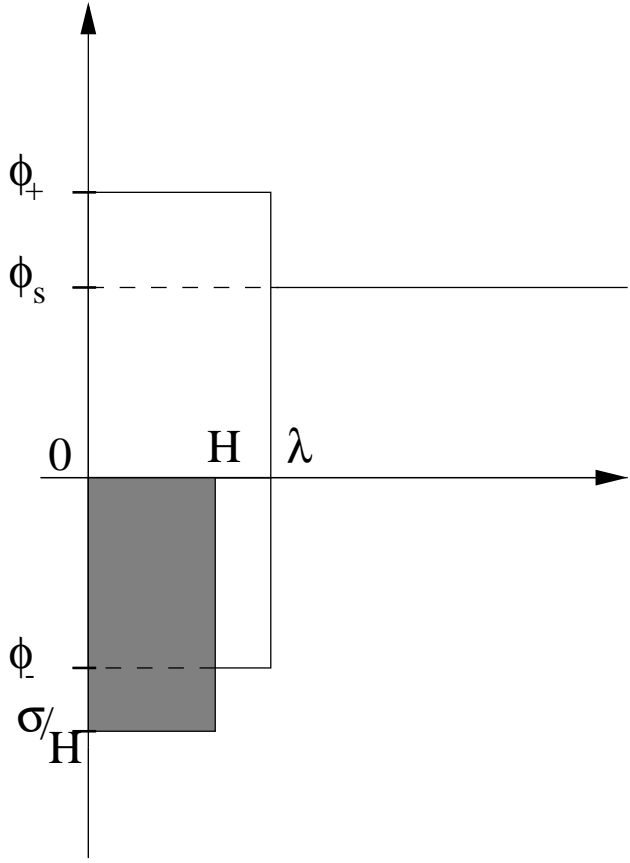,width=3.25in}}
\caption{The concentration profiles of ions within the box model for the 
diffuse layer. The uniformly smeared charge of the $p$ and $pt$ chains
is depicted by the shaded step function. It causes the concentration of
negative and positive ions, $\phi_{-}$ and $\phi_{+}$, within the proximal
layer of thickness $\lambda$, to deviate from the bulk value $\phi_s$.}
\end{figure}

The surface charge affects the distribution of ions within a proximal layer
of height $\lambda > H$, adjacent to the surface. Within this layer $n_{\pm}$ 
is the total number of univalent positive (negative) ions and $\phi_{\pm}=
n_{\pm}/\lambda A$ are the corresponding number concentrations. The electrical
potential in the ``box'', $\Psi$, determines the deviation of
$\phi_{\pm}$ from the bulk number concentration $\phi_{s}$ via $\phi_{\pm}
=\phi_{s}\exp(\pm e\Psi/kT)$ thus leading to the Donnan equilibrium
\begin{equation}
\phi_{+}\phi_{-}=\phi_{s}^{2}. \label{AI1}
\end{equation}
The overall electroneutrality of the proximal layer, $n_{+}-n_{-}=Q$ leads to
\begin{equation}
\Delta\phi=\phi_{+}-\phi_{-}=\sigma/\lambda.  \label{AI2}
\end{equation}
$\lambda$ is the ``neutralization length'' of the system in that the net 
charge of a thicker layer is zero and at higher altitude $\Psi = 0$.
Combining Eq.~\ref{AI1} and Eq.~\ref{AI2} leads to a quadratic equation 
$\phi_{+}^{2}-\frac{\sigma}{\lambda}\phi_{+}-\phi_{s}=0$ determining
$\phi_{\pm}$. Upon introducing the parameters $s=r_{D}/\Lambda$ and
$x=\lambda/\Lambda$ we obtain
\begin{equation}
\phi_{\pm} = \phi_{s} \left[ \pm \frac{2s^{2}}{x} + \left( 1 + 
\frac{4s^4}{x^2} \right)^{1/2} \right].  \label{AI3}
\end{equation}
The excess entropy of the ions in the box, with respect to the bulk, is
specified by $-S/k=n_{-}\ln(\phi_{-}/\phi_{s})+n_{+}\ln(\phi_{+}/\phi_{s})$.
Invoking Eq.~\ref{AI1} and Eq.~\ref{AI2} leads to $-S/k=A\sigma\ln(\phi_{+}
/\phi_{s})$ and the excess entropy per unit area is thus
\begin{equation}
-\frac{S}{Ak} = \sigma \ln \left[ \frac{2s^2}{x} + \left( 1 + 
\frac{4s^4}{x^2} \right)^{1/2} \right].  \label{AI4}
\end{equation}
The charge per unit area that is bound by a surface of height $z$ is
$ez(-\rho+\Delta\phi)$ when $0\leq z\leq H$ and $e(-\rho+z\Delta \phi)$ when
$H\leq z\leq \lambda$ (Figure 6). Consequently, the electrostatic field,
$E(z)$, as determined by the Gauss theorem, is
\begin{equation}
E(z)=\left\{
\begin{array}[c]{c}
E_{in}(z)=\frac{4\pi e\sigma}{\epsilon}\left(-\frac{1}{H}+
\frac{1}{\lambda}\right)  z \qquad 0 \leq z \leq H\\
E_{out}(z)=\frac{4\pi e\sigma}{\epsilon}\left(-1+\frac{z}{\lambda}
\right) \qquad H \leq z \leq \lambda
\end{array}
\right. \label{AI5}
\end{equation}
In the $H=0$ case the charge per unit area below $z$ is $e(-\rho+z\Delta \phi)$
and $E(z)=E_{out}(z)$ for $0\leq z\leq \lambda$. The associated electrostatic
energy per unit area, $W=\frac{\epsilon}{8\pi}\int_{0}^{\lambda} E^{2}(z)dz$ is
\begin{equation}
\frac{W}{kT}=\frac{\sigma x}{3}\left(1-\frac{H}{x\Lambda}\right)^{2}. \label{AI6}
\end{equation}
In the case of $H=0$ this reduces to $\sigma x/3$. Altogether, the
electrostatic free energy per unit area is
\begin{equation}
\frac{\gamma_{el}}{kT}=\frac{\sigma x}{3}\left(1-\frac{H}{x\Lambda}\right)^{2}
+\sigma \ln \left[\frac{2s^2}{x} + \left(1+\frac{4s^4}{x^2}\right)^{1/2}\right].
\label{AI7}
\end{equation}
The equilibrium condition $\partial \gamma_{el}/\partial x=0$ leads to
\begin{equation}
x^{2} \left( 1 + \frac{4s^4}{x^2} \right)^{1/2} \left[ 1- \left(
\frac{H}{x\Lambda}\right)^{2} \right]=6s^{2}. \label{AI8}
\end{equation}
We first consider the $H=0$ case when
\begin{equation}
x^{2} \left(1+\frac{4s^4}{x^2} \right)^{1/2}=6s^{2}. \label{AI9}
\end{equation}
In the high salt limit, when $s \ll 1$ this leads to equilibrium values of
$x \approx 6^{1/2} \, s$ and $\gamma_{el}/kT \approx 2 (2/3)^{1/2} \sigma s 
\approx 1.6 \, \sigma s$ or
\begin{equation}
\lambda \approx 6^{1/2} \, r_{D} \qquad \gamma_{el}/kT \approx 4 \pi (2/3)^{1/2}  
\, \sigma^{2} l_{B} r_{D} \label{AI10}
\end{equation}
as  compared to $\gamma_{el}/kT=\sigma s$ obtained from the rigorous solution
of the PB equation. In the opposite limit, of $s \gg 1$ corresponding to low
salt, Eq.~\ref{AI9} leads to $x \approx 3$ and $\gamma_{el}/kT \approx 2 \sigma
[\ln2s+(1-\ln 3)/2]$ or
\begin{equation}
\lambda \approx 3 \Lambda \qquad \gamma_{el}/kT \approx 2 \sigma \ln (4 \pi
\sigma l_{B}r_{D})  \label{AI11}
\end{equation}
while the rigorous solution of the PB equation is $\gamma_{el}/kT
\approx 2 \sigma [\ln 2s-1]$. Thus, the box model for the planar 
layer recovers the rigorous solutions of the PB equation up to 
numerical corrections. In the low salt regime it yields the correct 
leading term $\gamma_{el}/kT \approx 2 \sigma \ln s$. However, at 
high salt the box model overestimates $\gamma_{el}$ by $60\%$. 
This performance is indicative of the errors expected from the 
model for the diffuse layer.

When $H>0$ the equilibrium condition Eq.~\ref{AI8} is applicable. 
This equation differs from Eq.~\ref{AI9} in two respects: 
(i) a $\left[1-(H/x \Lambda)^{2}\right]$ factor arising from the 
modification of the charge distribution and the associated 
electrostatic energy and (ii) the problem now contains an 
additional length scale, $H$. We expect that $\lambda \gtrsim H$ 
and consequently the magnitude of $4 s^{4}/x^{2}= 4 r_{D}^{4}/
\lambda^{2}\Lambda^{2}$ can be large (small) even when $s=r_{D}/
\Lambda \ll 1$ ($s \gg 1$) provided $H \ll r_{D}$ ($H \gg r_{D}$). 
To allow for this last feature it is convenient to express 
Eq.~\ref{AI8} in terms of $y=\lambda/H$ instead of $x$ leading to
\begin{equation}
\left(y^{2}-1\right) \left[ 1 + \frac{4s^{4}}{y^{2}} \left(\frac{\Lambda}{H}
\right)^{2} \right]^{1/2} =6s^{2} \left(\frac{\Lambda}{H}\right)^{2}. 
\label{AII5}
\end{equation}
In analyzing the asymptotic solutions of this equation it is useful to
compare the neutralization length, $\lambda$, with $H$. Two principle regimes 
emerge. When $\lambda \gg H$ ($y \gg 1$), the structure of the diffuse layer
is irrelevant and we recover the solutions of the PB equation describing a 
charged planar layer. In this ``PB limit'' Eq.~\ref{AII5} reduces to
$y^{2} \left[ 1 + \frac{4s^4}{y^2} \left( \frac{\Lambda}{H} \right)^{2} 
\right]^{1/2} = 6s^{2} \left(\frac{\Lambda}{H}\right)^{2}$. Here we can 
again distinguish between two regimes. When $s^2 \Lambda/y H \gg 1$ 
this leads to $y \approx 3 \Lambda/H \gg 1$ while for $s^2 \Lambda/yH \ll 1$ 
we obtain $y \approx 6^{1/2} \, s\Lambda/H$. Altogether
\begin{equation}
\lambda \approx \left\{
\begin{array}[c]{lll}
3\Lambda & &H \ll \Lambda \text{ and } r_{D} \gg \Lambda\\
6^{1/2} \, r_{D} & & H \ll r_{D} \text{ and } r_{D} \ll \Lambda
\end{array}
\right. \label{AII8}
\end{equation}
When $\lambda \approx \Lambda$ the screening of the electrostatic potential
is due to the counterions of the charged layer. The coions, originating 
from the salt, dominate the screening when $\lambda \approx r_D$. 
The crossover between the ``salt screening'' (PBss) and ``counterions 
screening'' (PBcs) regimes in the PB limit occurs at $s^2 \Lambda/y H = 1$ 
leading to $s=1$ or $\Lambda=r_{D}$.

\begin{figure}
\centerline{\epsfig{file=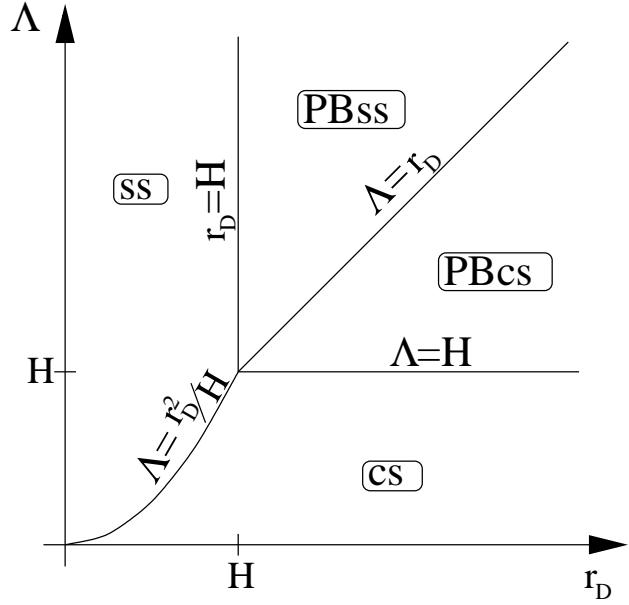,width=3.25in}}
\caption{The asymptotic regimes of the diffuse layer within the box model.
In the two PB regimes (PBcs and PBss) the neutralization length $\lambda$ 
is large, $\lambda \gg H$, and the layer behaves as a charged planar 
surface. In the two remaining regimes, $\lambda \gtrsim H$ and the charge 
distribution of the layer, $\rho$, plays a role. In the cs regions the
screening is dominated by the counterions while in the ss regions it is 
due to coions originating from the salt.}
\end{figure}

When $y \gtrsim 1$ the charge distribution within the diffuse layer plays 
an important role. In this case it is useful to express $y$ as $y=1+\delta$ 
and to solve with respect to $\delta \ll 1$. Eq.~\ref{AI8} reduces to $2 \,
\delta [ 1 + (2 \, s^2 \Lambda/H)^2 ]^{1/2} = 6 \, s^{2} \Lambda^2/H^2$. 
Consequently we can distinguish between two cases depending on the magnitude 
of $s^2 \Lambda/H$. When $s^{2}\Lambda/H \gg 1$ or $r_{D}^{2} \gg \Lambda H$ 
we obtain $\delta \approx 3 \Lambda/2H$. In the opposite limit, of $s^{2}
\Lambda/H \ll 1$ or $r_{D}^{2} \ll \Lambda H$ we obtain $\delta \approx 3 \,
(\Lambda/H)^{2}s^{2}$. That is
\begin{equation}
\lambda \approx \left\{
\begin{array}[c]{lll}
H+\frac{3}{2}\Lambda & & H \gg \Lambda \text{ and } r_{D}^{2} \gg \Lambda H\\
H+3\frac{r_{D}^{2}}{H} & & H \gg r_{D} \text{ and } r_{D}^{2} \ll \Lambda H
\end{array}
\right. \label{AII6}
\end{equation}
When $\lambda \approx H + 3 \Lambda/2$ the screening is due to the counterions
while for $\lambda \approx H + 3 r_D^2 /H$ it is dominated by the coions. 
The crossover between the ``salt screening'' (ss) and ``counterions screening''
(cs) regimes is specified by $s^{2}\Lambda/H=1$ or $\Lambda=r_{D}^{2}/H$.
Additional crossovers clearly occurs at $r_{D}=H$ and at $\Lambda=H$ (Figure 7). 

To obtain the corresponding asymptotic expressions for $\gamma_{el}$ it is 
convenient to rewrite Eq.~\ref{AI7} in terms of $y$ as
\begin{equation}
\frac{\gamma_{el}}{kT}=\sigma \left\{ \frac{H}{3\Lambda} \frac{(y-1)^{2}}{y} 
+ \ln \left[ \frac{2s^{2} \Lambda}{y H} + \left( 1 + \left(\frac{2 s^2 \Lambda}{yH}
\right)^{2}\right)^{1/2} \right] \right\}.  \label{AII10}
\end{equation}
When $\Lambda/H \ll 1$ and $s^{2}\Lambda/H \gg 1$ (cs regime), $y \approx 1 +
3\Lambda/2H$, the logarithmic term is dominant and $\gamma_{el}/kT \approx 
\sigma \ln (4s^{2}\Lambda/H)$. In the limit of $r_{D}/H \ll 1$ and $s^2 \Lambda/H 
\ll 1$ (ss regime), when $y \approx 1 + 3 r_{D}^{2}/H^{2}$, the logarithmic 
term can be expanded in powers of $s^{2}\Lambda/H$ leading to $\gamma_{el}/kT 
\approx 2 \sigma r_{D}^{2}/H \Lambda$. When $\Lambda/H \gg 1$ and $s^{2}\Lambda/y 
H \gg 1$ (PBcs regime), the logarithmic term is dominant and $\gamma_{el}/kT \approx 
\sigma [1+\ln (4s^{2}/3)]$ while for $r_{D}/H \gg 1$ and $s^{2}\Lambda/y H \ll 1$ 
(PBss regime), the logarithm can be expanded leading to $\gamma_{el}/kT \approx 
2 \, (2/3)^{1/2} \, \sigma r_{D}/\Lambda$. The four scaling regimes are 
summarized in Table I.
\begin{center}
\begin{equation*}
\begin{tabular}
[c]{ccccccc}
\ \ regime \ \ & \ \ \  & $\frac{\gamma_{el}}{kT}$ & \ \ \  & $\lambda$ 
& \ \ \ & range \\
\hline
cs & & $\sigma \ln \left(\sigma l_{B}r_{D}^{2}/H \right)$ & &
$H + 3 \Lambda/2$ & & $ \Lambda < H$ and $r_{D} > (\Lambda H)^{1/2}$ \\
\hline
ss & & $\sigma^{2}l_{B}r_{D}^{2}/H$ & & $H + 3 r_D^2/H$ & & $r_{D} < H$ 
and $r_{D}< (\Lambda H)^{1/2}$\\
\hline
PBcs & & $\sigma \ln(\sigma l_B r_D)$ & & $3\Lambda$ & & $\Lambda > H$ and 
$r_D > \Lambda$ \\
\hline
PBss & & $\sigma^{2}l_{B}r_{D}$ & & $6^{1/2} r_{D}$ & & $r_{D}> H$ and 
$r_{D}<\Lambda$ \\
\hline
\end{tabular}
\end{equation*}
\end{center}

\section{Appendix B: The Hybridization Isotherm at Low Salt.}

A novel form of the hybridization isotherm is obtained at low salt, 
when the screening is dominated by the counterions of the $p$ and $pt$ 
chains. This is the case when the concentration of counterions within 
the probe layer is much larger than the concentration of coions 
contributed by the salt leading to $r_{D} > (\Lambda H)^{1/2}$ and 
$\Lambda \ll H$. In this situation
\begin{equation}
\frac{\gamma_{el}}{kT}=\sigma \ln \left(8\pi\sigma l_{B}\frac{r_{D}^{2}}
{H}\right).  \label{III3}
\end{equation}
The hybridization isotherm in this ``counterion screening'' (cs) regime is
\begin{equation}
\frac{x}{c_{t}(1-x)}=K_{t}\exp\left[-\Gamma_{cs}-N\ln(1+x)\right]
\label{III4i}
\end{equation}
where $\frac{N}{kT}\frac{\partial \gamma_{el}}{\partial \sigma}\approx
\Gamma_{cs}+N\ln(1+x)$ and $\Gamma_{cs}=N[\ln\left(8\pi\sigma_{0}l_{B}
\frac{r_{D}^{2}}{H}\right)+1]$. The cs regime is of interest in that 
it provides an additional test for the diffuse layer model.

The authors benefitted from instructive discussions with T. Livache and P. Pincus. 
EBZ was funded by the CNRS and the Universit\'e Joseph Fourier.

\section*{References}


Bhanot, G., Y. Louzoun, J. Zhu, and C. DeLisi. 2003. 
The importance of thermodynamic equilibrium for high throughput gene 
expression arrays. \emph{Biophys. J.} 84:124-135.

Bloomfield, V.A., D.M. Crothers, and I. Tinoco. 2000.
Nucleic acids: structures, properties and functions. University science books,
Sausalito USA.

Borisov, O.V., E.B. Zhulina, and T.M. Birshtein. 1994.
Diagram of the states of a grafted polyelectrolyte layer.
\emph{Macromolecules} 27:4795-4803.

Cantor, C.R., and D.M. Schimmell. 1980. Biophysical
Chemistry. WH Freeman, New York

Chan, V., D.J. Graves, and S. McKenzie. 1995. The
biophysics of DNA hybridization with immobilized oligonucleotide probes.
\emph{Biophys. J.} 69:2243-2255.

Crozier, P.S., and M.J. Stevens. 2003. Simulations of 
single grafted polyelectrolytes chains: ssDNA and dsDNA. \emph{J. Chem. 
Phys.} 118:3855-3860.

Ekins, R., and P. Edwards. 1997. On the meaning of
``sensitivity''. \emph{Clin. Chem.} 43:1824-1830.

Evans, D.F., and H. Wennerstr\"{o}m. 1994. The Colloid
Domain. VCH, New York.

Frank-Kamenetskii, M.D., V.V. Anshelevich, and A.V.  Lukashin.
1987. Polyelectrolyte model of DNA. \emph{Sov. Phys. Usp.} 30:317-330.

Gerhold, D., T. Rushmore, and T. Caskey. 1999. DNA
chips: promising toys have become powerful tools. \emph{TiBS} 24:168-173. 

Graves, D.J. 1999. Powerful tools for genetic
analysis come of age. \emph{Trends Biotechnol.} 17:127-134.

Jancovici, B. 1982. Classical coulomb systems near
a plane wall. I \emph{J. Stat. Phys.} 28:43-65

Korolev, N., A.P. Lyubartsev, and L. Nordenski\"{o}ld. 
1998. Application of polyelectrolyte theories for analysis of DNA melting 
in the presence of $Na^+$ and $Mg^{2+}$ ions. \emph{Biophys. J.} 75:3041-3056.

Lakowicz, J. R. 1999. Principles of fluorescence spectroscopy.
Kluwer Academic, Plenum. 

Lau, A.W.C., D.B. Lukatsky, P. Pincus, and S.A. Safran. 2002. 
Charge fluctuations and counterion condensation. \emph{Phys. Rev. E} 
65:051502 (7 pages).

Levicky, R.T.M., T.M. Herne, M.J. Tarlov, and S.K. Satija.
Using self-assembly to control the structure of DNA monolayers on gold:
A neutron reflectivity study. 1998. \emph{J. Am. Chem. Soc.} 120:9787-9792.

Li, F. and G.D. Stormo. Selection of optimal DNA oligos for gene
expression arrays. 2001. \emph{Bioinformatics} 17:1067-1076.

Livshits, M.A., and A.D. Mirzabekov. 1996.
Theoretical analysis of the kinetics of DNA hybridization with
gel-immobilized oligonucleotides. \emph{Biophys. J.} 71:2795-2801.

Lockhart, D.J., et al. Expression monitoring by hybridization to
the high-density oligonucleotide arrays. 1996. \emph{Nat. Biotechnol.}
14: 1675-1680.

Lopez-Crapez, E., T. Livache, J. Marchand, and J. Grenier. 2001.
K-ras mutation detection by hybridization to a polypyrrole DNA chip.
\emph{Clin. Chem.} 47:186-194.

Marshall, A., and J. Hodgson. 1998. DNA chips: an
array of possibilities. \emph{Nat. Biotechnol.} 16:27-31.

Moore, W.J. 1972. Physical Chemistry. Longman, London.

Nelson, B.P., T.E. Grimsrud, M.R. Liles, R.M. Goodman, and 
R.M. Corn. 2001. Surface plasmon resonance imaging measurements of DNA and RNA 
hybridization adsorption onto DNA microarrays. \emph{Anal. Chem.} 73:1-7.

Niemeyer, C.M., and  D. Blohm. 1999. DNA microarrays.
\emph{Angew. Chem. Int. Ed.} 38:2865-2869.

Ohshima, H., and T. Kondo. 1993. Electrostatic
interactions of an ion penetrable sphere with a hard plate: contribution 
of image charges. \emph{J. Colloid Interface Sci.} 157:504-508.

Pardue, H.L. 1997. The inseparable triangle: analytical 
sensitivity, measurement uncertainty and quantitative resolution. 
\emph{Clin. Chem.} 43:1831-1837.

Peterson, A.W., R.J. Heaton, and R.M. Georgiadis.
2001. The effect of surface probe density on DNA hybridization.
\emph{Nucleic Acids Res.} 29:5163-5168.

Peterson, A.W., L.K. Wolf, and R.M. Georgiadis.
2002. Hybridization of mismatched and partially matched DNA at
surfaces. \emph{J. Am. Chem. Soc.} 124:14601-14607.

Pincus, P. 1991. Colloid stabilization with grafted polyelectrolytes. 
\emph{Macromolecules} 24:2912-2919.

Pirrung, M.C. 2002. How to make a DNA chip? \emph{Angew.
Chem. Int. Ed.} 41:1277-1289.

Sips, R. 1948. On the structure of a catalyst surface.
\emph{J. Chem. Phys.} 16:490-495.

Southern, E., K. Mir, and M. Shchepinov. 1999.
Molecular interactions on microarrays. \emph{Nat. Genet.} 21:5-9.

Steel, A.B., R.L. Levicky, T.M. Herne, and M.J. Tralov. 2000. 
Immobilization of nucleic acids at solid surfaces: effect of oligonucleotide 
length on layer assembly. \emph{Biophys. J.} 79:975-981.

Tibanyenda, N., S.H. De Bruin, C.A. Haasnoot, G.A. van der Marel, J.H. van Boom, 
and C.W. Hilbers. 1984. The effect of single base-pair mismatches on the duplex 
stability of d(T-A-T-T-A-A-T-A-T-C-A-A-G-T-T-G).d(C-A-A-C-T-T-G-A-T-A-T-T-A-A-T-A).
\emph{Eur. J. Biochem.} 139:19-27.

Vainrub, A., and M.B. Pettitt. 2002. Coulomb
blockage of hybridization in two-dimensional DNA arrays. 
\emph{Phys. Rev. E} 66:041905 (4 pages).

Vainrub, A., and M.B. Pettitt. 2003. Surface
electrostatic effects in oligonucleotide microarrays: control and 
optimization of binding thermodynamics. \emph{Biopolymers}, 68:265-270.

Wang, J. 2000. From DNA biosensors to gene chips.
\emph{Nucleic Acids Res} 28:3011-3016.

Wittmer, J., and J.F. Joanny. 1993. Charged diblock copolymers at interfaces.
\emph{Macromolecules} 26:2691-2697.


\end{document}